\DocumentMetadata{}
\documentclass[sigconf]{acmart}

\usepackage{array}

\usepackage{xspace}
\usepackage{colortbl}
\usepackage{xspace}
\usepackage{hyperref}

\AtBeginDocument{%
  }

\setcopyright{acmlicensed}
\copyrightyear{2018}
\acmYear{2018}
\acmDOI{XXXXXXX.XXXXXXX}

\acmConference[Conference acronym 'XX]{Make sure to enter the correct
  conference title from your rights confirmation email}{June 03--05,
  2018}{Woodstock, NY}

\acmISBN{978-1-4503-XXXX-X/2018/06}

\usepackage{xspace}
\usepackage{colortbl}
\usepackage{xspace}
\usepackage{enumitem}
\usepackage{soul}
\usepackage{dirtytalk}
\usepackage{epigraph} 
\usepackage{multirow}
\usepackage{hyperref}

\usepackage{xcolor,colortbl}
\definecolor{lblue}{HTML}{135ABE}

\usepackage{ifthen}
\usepackage{xcolor}

\copyrightyear{2026}
\acmYear{2026}
\setcopyright{cc}
\setcctype{by}
\acmConference[CHI EA '26]{Extended Abstracts of the 2026 CHI Conference on Human Factors in Computing Systems}{April 13--17, 2026}{Barcelona, Spain}
\acmBooktitle{Extended Abstracts of the 2026 CHI Conference on Human Factors in Computing Systems (CHI EA '26), April 13--17, 2026, Barcelona, Spain}
\acmPrice{}
\acmDOI{10.1145/3772363.3778710}
\acmISBN{979-8-4007-2281-3/2026/04}

\begin{document}



%
%
\title{Human-AI Interaction Alignment: Designing, Evaluating, and Evolving Value-Centered AI For Reciprocal Human-AI Futures}

\renewcommand{\shorttitle}{Human-AI Alignment for Reciprocal Human-AI Futures}






\author{Hua Shen}
\orcid{1234-5678-9012}
\affiliation{%
  \institution{NYU Shanghai, New York University}
  \country{China}
}
\email{huashen@nyu.edu}

\author{Tiffany Knearem}
\orcid{0000-0003-4928-6225}
\affiliation{%
  \institution{MBZUAI}
  \country{United Arab Emirates}
}
\email{tiffany.knearem@mbzuai.ae.ac}

\author{Divy Thakkar}
\orcid{}
\affiliation{%
  \institution{Google}
  \country{United States}
}
\email{dthakkar@google.com}

\author{Pat Pataranutaporn}
\orcid{}
\affiliation{%
  \institution{Massachusetts Institute of Technologyy}
  \country{United States}
}
\email{patpat@media.mit.edu}

\author{Anoop Sinha}
\orcid{}
\affiliation{%
  \institution{Google, Paradigms of Intelligence}
  \country{United States}
}
\email{anoopsinha@google.com}

\author{Yike Shi}
\orcid{}
\affiliation{%
  \institution{Carnegie Mellon University, New York University}
  \country{United States}
}
\email{yikes@andrew.cmu.edu}

\author{Jenny Liang}
\orcid{}
\affiliation{%
  \institution{Carnegie Mellon University}
  \country{United States}
}
\email{jtliang@andrew.cmu.edu}

\author{Lama Ahmad}
\orcid{}
\affiliation{%
  \institution{OpenAI}
  \country{United States}
}
\email{lama@openai.com}

\author{Tanu Mitra}
\orcid{}
\affiliation{%
  \institution{University of Washington}
  \country{United States}
}
\email{tmitra@uw.edu}

\author{Brad A. Myers}
\orcid{}
\affiliation{%
  \institution{Carnegie Mellon University}
  \country{United States}
}
\email{bam@cs.cmu.edu}

\author{Yang Li}
\orcid{}
\affiliation{%
  \institution{Google DeepMind}
  \country{United States}
}
\email{liyang@google.com}

\renewcommand{\shortauthors}{}

\begin{abstract}
The rapid integration of generative AI into everyday life underscores the need to move beyond unidirectional alignment models that only adapt AI to human values. This workshop focuses on bidirectional human-AI alignment, a dynamic, reciprocal process where humans and AI co-adapt through interaction, evaluation, and value-centered design. Building on our past CHI 2025 BiAlign SIG and ICLR 2025 Workshop, this workshop will bring together interdisciplinary researchers from HCI, AI, social sciences and more domains to advance value-centered AI and reciprocal human-AI collaboration. We focus on embedding human and societal values into alignment research, emphasizing not only steering AI toward human values but also enabling humans to critically engage with and evolve alongside AI systems. Through talks, interdisciplinary discussions, and collaborative activities, participants will explore methods for interactive alignment, frameworks for societal impact evaluation, and strategies for alignment in dynamic contexts. This workshop aims to bridge the disciplines' gaps and establish a shared agenda for responsible, reciprocal human-AI futures.

\end{abstract}



\maketitle

\section{Background and Motivation}
Rapid advances in general-purpose and generative AI have precipitated the urgent need to align these systems with the values, ethical principles, and goals of individuals and society at large~\cite{maity2025healthcarellms,xu2024llmedusurvey,ma2024teach,vaananen2021ai,prabhudesai2025here}. This need, commonly referred to as AI alignment~\cite{wiki_AI_alignment,terry2023ai}, is crucial to ensure that AI systems function in ways that are not only effective but also consistent with human values, minimizing harm and maximizing societal benefits~\cite{goyal2024designing,ouyang2022training,santurkar2023whose,scheurer2023large,carroll2023characterizing,jakesch2023co}. Traditionally, alignment has been viewed as a static, one-way process, with AI shaped to achieve desired outcomes and avoid negative side effects. Yet, as AI systems increasingly permeate daily life and take on complex decision-making roles, this unidirectional approach proves inadequate~\cite{carroll2024ai}. AI systems and humans interact in evolving and unpredictable ways, generating feedback loops that influence both AI behavior and human responses \cite{pataranutaporn2023influencing, liu2025heterogeneous, fang2025ai, pataranutaporn2024cyborg}. This dynamic relationship necessitates a shift toward \emph{bidirectional human-AI alignment} -- a paradigm that treats alignment as a continuous, reciprocal process of mutual adaptation~\cite{shen2024towards}. From this perspective, alignment requires not only steering AI toward human goals but also empowering humans to critically engage with, recalibrate, and evolve alongside AI systems.

\textbf{Past Workshops and Community Interest: } 
We have observed a growing number of workshops advancing AI alignment within the AI-centered research community, such as the NeurIPS 2024 Pluralistic Alignment Workshop, ICML 2024 Human Feedback for AI Alignment Workshop, and ICLR 2024 Representation Alignment Workshop. Yet, the \textbf{voice of the HCI community—bringing a human-centered perspective to alignment—has been notably absent}. To help address this gap, our team has begun building this vision through the ICLR 2025 Bidirectional Human-AI Alignment Workshop~\cite{shen2025iclr} and the CHI 2025 Bidirectional Human-AI Alignment SIG~\cite{shen2025bidirectional}. Both efforts attracted overwhelming interest: the ICLR workshop received more than 100 submissions, and the CHI SIG drew over 100 participants, filling the room to capacity. Many additional attendees who could not join requested access to materials, underscoring the strong demand for an interdisciplinary forum on this topic. The success of these past events in 2025 \textbf{highlights both the urgency of this research area and the opportunity for CHI to host a more expansive, dedicated workshop} where researchers can share insights, methods, and perspectives.

\textbf{Novel Contributions of CHI 2026 Workshop: } 
Building on this momentum, this CHI 2026 workshop -- serving as the HCI home of our 2nd BiAlign Workshop -- introduces several new initiatives to deepen engagement and expand the community: (1) \emph{Interdisciplinary Integration}: Bridge interdisciplinary research via structured sessions and collaborations; (2) \emph{Interactive Knowledge Creation}: Group activities such as collaborative paper prototyping, solution ideation, and concept mapping to co-develop new ideas; (3) \emph{Expanded Accessibility}: A hybrid format supported by pre-recorded talks, shared materials, and a dedicated Slack channel to engage participants worldwide; (4) \emph{Sustained Community Building}: Post-workshop initiatives including an open repository, ongoing discussion spaces, and opportunities for collective publications.
Through these contributions, the workshop will advance value-centered approaches to bidirectional human-AI alignment, positioning CHI as a central venue for shaping reciprocal human-AI futures grounded in responsibility, values, and collaboration.

    


\section{Workshop Goals and Themes}

The overarching goal of this workshop is to establish a sustained, interdisciplinary forum for advancing bidirectional human-AI alignment — a paradigm that emphasizes dynamic, reciprocal processes of co-adaptation between humans and AI systems, grounded in human and societal values. Specifically, the workshop goals include:

\begin{enumerate}[topsep=0pt, labelwidth=*,leftmargin=1.8em,align=left, labelsep=5pt,label=G.\arabic*]

\item \textbf{Operationalize Human and Societal Values: }  
We will identify and discuss frameworks for translating abstract values—such as fairness, agency, and responsibility—into actionable design principles and technical requirements. The workshop will surface practical strategies for embedding these values into the development and deployment of AI systems.


\item \textbf{Advance Design and Interaction Mechanisms: } 
Drawing from Human–Computer Interaction (HCI) methods, we will explore how interaction design, user experience research, and participatory approaches can enable humans to shape, critique, and recalibrate generative AI systems in real time. A particular emphasis will be placed on techniques that empower diverse stakeholders to meaningfully engage with, and guide, AI behavior.

\item \textbf{Explore Dynamic Evaluation Approaches: } The workshop will examine approaches to evaluating alignment at individual, community, and societal levels—balancing technical performance with societal impact. Participants will also consider strategies for sustaining alignment over time, recognizing that both humans and AI systems evolve as models acquire more advanced reasoning and adaptive capabilities.


\item \textbf{Foster Interdisciplinary Collaboration and Build Community: } 
Create a forum for researchers and practitioners in HCI, AI, and the social sciences to exchange perspectives, bridge disciplinary gaps, and shape shared research agendas. Through networking and collaborative activities, the workshop will strengthen the BiAlign community and lay the groundwork for sustained engagement beyond the event.



\end{enumerate}

%
\textbf{Themes: }
We will structure the workshop around four interrelated themes. Each theme will be introduced through short talks and exemplars, followed by interactive discussions and collaborative activities: 
\begin{itemize}[topsep=0pt, labelwidth=*,leftmargin=1.8em,align=left]
    \item \textbf{Value-Centered Alignment Objectives: } Explores which human and societal values should guide reciprocal human-AI alignment and how these values can be articulated and translated into technical and design processes. 
    \begin{itemize}
        \item \emph{Research Questions: } What fundamental human and societal values should guide reciprocal human-AI alignment?
        %
        %
        In what ways might HCI contribute to the articulation and translation of values into technical and design processes?
    \end{itemize}
    \begin{itemize}
        \item \emph{Keywords \& Example Papers}: pluralistic values, human agency, cultural perspectives, value-sensitive design, etc~\cite{friedman1996value,shen2024valuecompass,shen2025mind}.
    \end{itemize}
    \item \textbf{Developing Interfaces and Interactions for Alignment: } Investigates design mechanisms—such as interfaces, interaction modalities, explanation systems, and participatory methods that empower humans to steer, critique, and co-create with AI systems.
    \begin{itemize}
        \item \emph{Research Questions}: What design mechanisms can help humans shape and steer AI behavior?
        %
        %
        What role do co-creation and participatory design methods play in aligning AI with evolving human needs?
        How do we uplift and retain human agency via effective human-AI collaboration?
    \end{itemize}
    \begin{itemize}
        \item \emph{Keywords \& Example Papers}: interactive alignment, UX for AI, participatory design, human-AI collaboration, etc~\cite{convxai,wu2023scattershot,gordon2022jury, danry2023don}.
    \end{itemize}
    \item \textbf{Evaluating Alignment and Societal Impacts: } Examines frameworks and methodologies for assessing bidirectional alignment, including both technical effectiveness and broader impacts such as trust and social well-being. 
    \begin{itemize}
        \item \emph{Research Questions:} 
        How should bidirectional alignment be measured—both technically and socially?
        What frameworks and methodologies can capture the broader impacts of alignment (e.g., trust, collective well-being, economic impact)?
        %
    \end{itemize}
    \begin{itemize}
        \item \emph{Keywords \& Example Papers}: alignment evaluation, societal impact, trust, responsible AI, etc~\cite{dammu2024they,kapoor2024societal,jakesch2023co}.
    \end{itemize}
    %
    %
    \item \textbf{Dynamic Co-Evolution of Human-AI Futures: } Considers alignment as an evolving process, highlighting strategies for sustaining reciprocal adaptation as both humans and AI change over time and across contexts.

    \begin{itemize}
        \item \emph{Research Questions}: 
        How have alignment goals and practices evolved over time, as humans and AI systems mutually adapted?
        %
        %
        How can we envision and design for long-term reciprocal futures of human-AI collaboration?
    \end{itemize}
    
    \begin{itemize}
        \item \emph{Keywords \& Example Papers}: adaptability, resilience, lifelong learning, co-evolution, etc~\cite{carroll2024ai,jiang2018beyond,pataranutaporn2023influencing,guo2024exploring}.
    \end{itemize}

\end{itemize}

Achieving these goals requires a diverse group of interdisciplinary researchers and practitioners, working together in open dialogue to shape, define and execute on alignment goals. 
This workshop aims to bring together experts from HCI, AI, psychology, social sciences, and more domains to advance interdisciplinary research and collaboration on bidirectional human-AI alignment.

\section{Organizers}

%
This workshop brings together an interdisciplinary team of organizers spanning academia and industry with global representations from the United States, China, and United Arab Emirates. The organizers contribute expertise across HCI, CSCW, NLP, ML, psychology, AI safety and governance. The team includes researchers who have shaped the study of bidirectional human-AI alignment and working persistently to contribute this research to broader interdisciplinary communities.
The organizers also bring substantial experience in organizing successful workshops and tutorials at premier venues such as CHI, CSCW, ICLR, EMNLP and more, ensuring effective facilitation and impactful outcomes.



\textbf{\href{https://hua-shen.org/}{Hua Shen, Ph.D.}} is an Assistant Professor of Computer Science at NYU Shanghai, New York University. Her work focuses on Bidirectional Human–AI Alignment: enabling humans to interactively explain, evaluate, and collaborate with AI systems, while integrating human feedback and values to improve AI systems.
She has been recognized as a 2023 Rising Star in Data Science, and received multiple awards, including AIED 2024 Best Paper and Best Interactive Event, CSCW 2023 Best Demo, and IUI 2023 Best Paper Honorable Mention, and Google Research Scholarships. 
She is serving as Associate Chair for CHI 2026-2025, CHI LBW 2024, Program Committees for ACL, EMNLP, and more. She initiated the 2025 BiAlign CHI SIG and ICLR workshop, NeurIPS 2025 Human-AI Alignment Tutorial, CSCW 2025 Design for Hope workshop, EMNLP 2025 WiNLP workshop for HCI and AI communities.

\textbf{\href{https://tknearem.wixsite.com/tknearem}{Tiffany Knearem, Ph.D.}} is an Affiliated Assistant Professor at the Mohamed bin Zayed University of Artificial Intelligence (MBZUAI) and the head of TK Research, a UX and HCI research consultancy. She holds a Ph.D. in information sciences and technology from Pennsylvania State University, advised by Prof. John M. Carroll. Her recent research interests span human-AI alignment, AI-supported design workflows, and community informatics. She co-organized the CHI 2024 and CHI 2025 Computational UI workshops, and the CHI 2025 Bi-Align SIG.

\textbf{\href{https://sites.google.com/view/divythakkar/home}{Divy Thakkar, Ph.D.}} is a Staff Program Manager and Researcher at Google DeepMind, where he is building new interactions and human-ai collaboration mechanisms for Gemini. His research has earned recognition at top HCI conferences, including CHI and CSCW. Thakkar completed his Ph.D. in Computer Science at City St. Georges, University of London.

\textbf{\href{https://www.media.mit.edu/people/patpat/overview/}{Pat Pataranutaporn, Ph.D.}} 
is an Assistant Professor at the MIT Media Lab, where he directs the Cyborg Psychology Research Group and co-directs the Advancing Humans with AI (AHA) Program. His research develops AI systems that foster human flourishing, including pioneering studies on generative AI for learning and self-development. His work has been published in Nature Machine Intelligence, ACM SIGCHI, and SIGGRAPH, PNAS, and featured in outlets such as The New York Times, Scientific American, and MIT Tech Review. Recognized by TIME’s Best Inventions of 2023 and Fast Company’s World Changing Ideas, his projects have been exhibited internationally and he has collaborated with NASA, OpenAI, Microsoft Research, and others. He also co-designed one of MIT’s first courses on Generative AI and co-created Netflix’s 2024 sci-fi anthology Tomorrow and I.

\textbf{\href{https://research.google/people/anoopsinha/}{Anoop Sinha, Ph.D.}} is currently a Research Director focusing on AI and Future Technologies at Google, where he leads research into new interfaces and previously directed cross-company AI initiatives involving data and development. Holding a Ph.D. from the University of California, Berkeley, his career spans significant leadership roles in AI and HCI across major tech companies, including Head of Siri ML and Knowledge at Apple and Sr. Applied Research Scientist Manager at Meta (FAIR X - HCI \& AI). His expertise lies at the intersection of machine learning, human-computer interaction, search quality, and knowledge representation. 

\textbf{\href{https://www.linkedin.com/in/yike-cassandra-shi}{Yike (Cassandra) Shi}} is a Research Associate jointly affiliated with New York University and Carnegie Mellon University. Her research focuses on requirements-driven prompting for LLMs, where she developed a web-based system that compiles user requirements into optimized prompts and integrates automated evaluation mechanisms. She also has industry experience as an AI Infrastructure Intern at DeepLang AI, where she optimized CUDA kernels and improved inference efficiency for quantized models. Shi’s academic projects span speech recognition, face recognition, and Transformer-based speech-to-text systems, as well as systems-level programming, database design, and game development. She has been recognized on the Dean’s List across multiple semesters.

\textbf{\href{https://www.media.mit.edu/people/patpat/overview/}{Jenny T. Liang}} is a PhD student in Software Engineering at Carnegie Mellon University, advised by Brad A. Myers. Her research sits at the intersection of software engineering, HCI, and applied machine learning, focusing on how developers interact with AI-powered tools and how to design more usable systems.
She has published in leading venues such as ICSE, FSE, and CHI, receiving awards including the ACM SIGSOFT Distinguished Paper Award. In addition to her research, Jenny has been active in community-building — organizing workshops at ICSE and CHI that bring together researchers across software engineering, HCI, and AI. She also has industry experience through internships at Apple, Microsoft, and AI2, and is dedicated to mentoring and service within the academic community.

\textbf{\href{https://asiafoundation.org/people/lama-ahmad/}{Lama Ahmad, Ph.D.}} 
is a researcher and technology professional currently leading partnerships and research on the risks and social impacts of AI at OpenAI’s Safety Systems team. She also serves as a Term Trustee for The Asia Foundation, guiding governance and strategy. 
Previously, Lama worked on Facebook’s Open Research \& Transparency team, focusing especially on democracy, elections, and the societal consequences of social media platforms. During her Luce Scholar year, she studied the ethics of data-driven technologies at the U.N. Global Pulse Lab in Jakarta, applying human-centered design across Southeast Asia. 
%
She is a passionate advocate for equity, inclusion, and interdisciplinary approaches in tech and policymaking.

\textbf{\href{https://faculty.washington.edu/tmitra/}{Tanu Mitra, Ph.D.}} is an Associate Professor in the Information School at the University of Washington, with an affiliate appointment in the Paul G. Allen School of Computer Science \& Engineering. She is also the Founding Co-Director of RAISE (Responsibility in AI Systems and Experiences). Her research focuses on Human-Centered AI and Responsible AI, combining computational techniques, NLP, and social science principles to study human behavior and interaction in large-scale online systems. An interdisciplinary scholar, Mitra employs methods from HCI, data science, and AI to design systems that foster responsible and effective human-computer and human-human communication. Prior to UW, she was an Assistant Professor in Computer Science at Virginia Tech and received her Ph.D. in Computer Science from Georgia Tech.

\textbf{\href{https://www.cs.cmu.edu/~bam/}{Brad A. Myers, Ph.D.}} is the Charles M. Geschke Director of the Human-Computer Interaction Institute and Professor in the School of Computer Science at Carnegie Mellon University, with an affiliated appointment in the Software and Societal Systems Department. He is an ACM Fellow, IEEE Life Fellow, CHI Academy member, and recipient of the 2017 ACM SIGCHI Lifetime Achievement Award in Research. His book, \textit{Pick, Click, Flick!} won the 2025 CBI HCI History Prize. Myers has authored or edited more than 550 publications, including three books, with 19 Best Paper Awards and 6 Most Influential Paper Awards. He has consulted for over 90 companies on UI design and regularly teaches HCI and software design. His research spans interaction techniques, developer experience, API usability, end-user software engineering, programming by example, and visual programming. He has helped organize and run multiple workshops and conferences.

\textbf{\href{https://yangli169.github.io/yangl.org/}{Yang Li, Ph.D.}} is a Senior Staff Research Scientist at Google DeepMind and Affiliate Associate Professor at the University of Washington CSE. His research lies at the intersection of HCI and AI, with a focus on user interface understanding, automation, generation, and code generation for UIs and apps. He has advanced deep learning methods such as Fourier Positional Encoding and area attention, and created influential benchmarks like screen2words and seq2act. Li pioneered on-device interactive ML on Android, leading to features like next app prediction and Gesture Search. He has published widely across HCI and ML venues (CHI, UIST, ICML, NeurIPS, ICLR, ACL, CVPR) and received multiple Best Paper and Lasting Impact Awards. An ACM Distinguished Scientist, Li co-edited AI for HCI: A Modern Approach and organized the inaugural AI \& HCI workshops at ICML. He earned his Ph.D. in Computer Science from the Chinese Academy of Sciences and completed postdoctoral research at UC Berkeley EECS.

\section{Workshop Schedule and Activities}

We propose a long, in-person workshop with 180 minutes (including breaks). The workshop is designed to \emph{balance knowledge sharing, interactive discussions, and collaborative activities}, allowing participants to meaningfully connect with others in the AI alignment community. The tentative workshop schedule is detailed in Table~\ref{tab:schedule}.

\begin{table*}[ht]
\centering
\begin{tabular}{@{}p{3cm} | p{7.5cm}@{}}
\toprule
\textbf{Time} & \textbf{Activity} \\
\midrule
\multicolumn{2}{l}{\textbf{Session 1 (90min)}} \\
\midrule
15 min & Welcome \& Overview \\
20 min & Keynote Talk 1: Lama Ahmad (OpenAI) \\
20 min & Lightning Talks by Authors \\
35 min & Group Activity 1: Concept Mapping \& Solution Ideation \\

\midrule
\multicolumn{2}{l}{\textbf{Session 2 (90min)}} \\
\midrule

20 min & Keynote Talk 2: Elizabeth F. Churchill (MBZUAI) \\
20 min & Poster Session \& Networking \\
30 min & Group Activity 2: On-the-spot Paper Writing \\
20 min & Insight Sharing \& Closing Remarks  \\ 
\bottomrule
\end{tabular}
\caption{Workshop schedule (180-min long session) with papers, posters, group activities, and discussions.}
\label{tab:schedule}
\end{table*}

\subsection{Before the Workshop}
Before the workshop session, we will invite participants through social media promotion and professional mailing lists. We expect attendees from diverse backgrounds with varying levels of familiarity and seniority with the topic.
To facilitate pre-workshop engagement, we previously created a \href{https://join.slack.com/t/bi-alignworkshops/shared_invite/zt-2vpuf45n7-zP8DcmoRwjqfCxVQ4f5_Kw}{BiAlign Slack channel} (now with 250+ participants), where attendees share materials, papers, and networking opportunities. For CHI 2026, we will expand these platforms to support both pre-event collaboration and post-workshop discussions.

\subsection{Workshop Schedule}
The workshop will be held for 180 min with breaks in the afternoon and is organized into two main sessions. We describe more details about the activities below.

\textbf{Keynote Talks: }
Two distinguished scholars anchor our program. \href{https://asiafoundation.org/people/lama-ahmad/}{Dr. Lama Ahmad (OpenAI)} will open the first session with a keynote that connects alignment research to human-centered values, setting the tone for the day. \href{https://elizabethchurchill.com/}{Dr. Elizabeth F. Churchill (MBZUAI)} will kick off the second session with a forward-looking talk, drawing from her expertise in HCI and AI to inspire cross-disciplinary dialogue. Each keynote will include space for questions, ensuring participants can engage directly with the speakers.

\textbf{Paper Presentations: } 
Accepted papers will be shared through \emph{lightning talks}, giving authors an opportunity to present their ideas in a lively, fast-moving format that sparks curiosity and discussion. A \emph{poster session} later in the workshop will encourage one-on-one exchanges, deeper conversations, and networking across disciplines, providing ample opportunities for participants to find shared interests.

\textbf{Group Activity 1 | Concept Mapping \& Solution Ideation: } 
The first collaborative session centers on a \emph{concept mapping and solution ideation} exercise. Working in groups, participants will identify challenges, connect ideas across disciplines, and co-develop creative solutions. This interactive mapping process ensures that every participant’s perspective contributes to a shared vision.

\textbf{Group Activity 2 | On-the-spot Paper Writing: } 
Building on the momentum, the second group session introduces an \emph{on-the-spot paper writing} challenge. Teams will synthesize earlier discussions into short outlines or position pieces, capturing fresh insights in a format that can lead to concrete post-workshop collaborations. This activity not only encourages creativity but also creates tangible outputs participants can take forward.

\textbf{Insight Sharing \& Closing Remarks: } 
The workshop concludes with \emph{Closing Remarks}, summarizing takeaways and outlining next steps for sustained engagement. 
Participants are encouraged to continue discussions and collaboration through the dedicated \emph{Slack Channel} and workshop website. 
This structure is designed to foster active participation, collaboration, and networking, while allowing participants to explore human-AI alignment topics in depth.

\subsection{Post-Workshop \& Plans to Publish Proceedings}
We plan to compile a comprehensive report summarizing key discussions, presentations, and findings, which will be shared via open-access platforms such as ArXiv and the workshop website. Outcomes from the On-the-spot Paper Writing session may be developed into full papers for submission to HCI venues such as CHI.

\textbf{Plans to Publish Proceedings: } We plan to publish workshop proceedings by collecting accepted papers and curating an edited volume, special journal issue (e.g., ACM ToCHI or ACM TIIS), or as online proceedings via platforms such as \href{https://ceur-ws.org}{CEUR-WS}. This will ensure the workshop outcomes reach a broad and sustained audience while supporting the continued development of the AI alignment research community.

\textbf{Offline Materials: }
Offline and asynchronous access will be provided through the workshop website and Slack, including the program schedule, list of organizers and speakers, pre-prints of accepted papers, and recordings of presentations, which will also be made publicly available on YouTube.

\subsection{Intended Community \& Expected Size}
We expect workshop attendees to include academic and industry researchers and practitioners broadly interested in AI alignment topics, coming from diverse disciplines such as human-computer interaction, AI, machine learning, psychology, and social sciences, without requiring deep technical expertise in AI. 

\textbf{Interdisciplinary Community Connections: }
To foster interdisciplinary community connections, we will also host a Machine Learning-oriented \href{https://openreview.net/forum?id=HcTiacDN8N}{Bidirectional Human-AI Alignment Workshop}, and our shared \href{https://join.slack.com/t/bi-alignworkshops/shared_invite/zt-2vpuf45n7-zP8DcmoRwjqfCxVQ4f5_Kw}{Slack Channel} enables ongoing collaboration among researchers from HCI and ML communities, and other alignment-focused communities. 
To ensure meaningful, in-depth discussions, the workshop is tailored for \textbf{30-50} in-person participants. We may adjust the structure to accommodate a slightly larger group while preserving interactive engagement.


\section{Logistics and Accessibility}
We are committed to creating an inclusive workshop environment for all participants, including those with cognitive, mental health, or physical disabilities. Authors will be encouraged to make their position papers accessible, and guidance will be provided for accepted papers, including alt-text for images and tables and clear document structure for screen readers. During the workshop, participants will be asked to follow accessibility best practices, such as using microphones and enabling captions for all presentations.

\section{Call For Participation}
We invite researchers and practitioners from academia and industry to join our Bidirectional Human-AI Alignment (BiAlign) Workshop at CHI 2026. As AI systems increasingly permeate everyday life, alignment requires dynamic, reciprocal processes in which humans and AI adapt to each other over time. This workshop provides an interactive forum to explore value-centered alignment, human-AI interaction design, evaluation methods, and strategies for dynamic co-evolution.
The workshop will feature paper presentations, poster sessions, and collaborative group activities such as on-the-spot paper writing, concept mapping, and solution ideation. These activities are designed to foster interdisciplinary knowledge creation, critical discussion, and co-development of new research directions.
We welcome submissions of position papers, posters, or brief research notes that address human-AI alignment from HCI, AI, psychology, social sciences, or related domains. Accepted participants are expected to attend the workshop, with at least one organizer per accepted submission present.
Key workshop topics include:

\begin{itemize}[topsep=0pt, labelwidth=*,leftmargin=1.8em,align=left, labelsep=5pt]
    \item \textbf{Value-Centered Alignment Objectives: }Embedding fairness, agency, equity, and responsibility into AI systems
    \item \textbf{Designing and Interacting for Alignment: }Interfaces, explanation, and participatory methods for steering AI
    \item \textbf{Evaluating Alignment and Societal Impacts: }Metrics and frameworks for technical and social assessment
    \item \textbf{Dynamic Co-Evolution of Human-AI Futures: } Strategies to maintain alignment as humans-AI mutually adapt
\end{itemize}
We expect 30-50 in-person participants, ensuring a highly interactive environment.
We welcome participants
from HCI, CSCW, AI, design, psychology, communication, and policy. Join us to connect with the growing BiAlign community, engage in hands-on collaborative activities, and help shape reciprocal and responsible human-AI futures.
%


\bibliographystyle{ACM-Reference-Format}
\bibliography{papers}


\end{document}